\documentclass{PoS}

\usepackage{graphicx}
\usepackage{amssymb} 
\usepackage{amsmath}
\usepackage[utf8]{inputenc}
\usepackage[T1]{fontenc}
\usepackage{multirow,bigdelim}
\usepackage{tabularx,ragged2e}
\usepackage{txfonts}

\usepackage{float} 
\usepackage[caption=false]{subfig} 

\usepackage[normalem]{ulem}

\newcommand{\sT}{{\scriptscriptstyle T}}
\renewcommand{\d}{\mathrm{d}}
\def\slash#1{\setbox0=\hbox{$#1$}               
        \dimen0=\wd0                            
        \setbox1=\hbox{/} \dimen1=\wd1          
        \ifdim\dimen0>\dimen1                   
        \rlap{\hbox to \dimen0{\hfil/\hfil}}    
        #1                                      
        \else
        \rlap{\hbox to \dimen1{\hfil$#1$\hfil}} 
        /                                       
        \fi}        

\newcommand{\nn}{\nonumber}

\def\GeV{{\rm GeV}}
\def\GeV2{{\rm GeV}^2}

\newcommand{\Q}{{\cal Q}}

\newcommand{\CS}{{\rm CS}}

\newcommand{\mc}[1]{\mathcal{#1}}
\newcommand{\kT}{\bm k_{\sT}}
\newcommand{\koneT}{\bm k_{1\sT}}
\newcommand{\ktwoT}{\bm k_{2\sT}}
\newcommand{\qT}{\boldsymbol{P}_{\Q\Q\sT}}

\renewcommand{\d}{\mathrm{d}}

\newcommand{\bm}[1]{\mbox{\boldmath $#1$}}

\let\OLDthebibliography\thebibliography
\renewcommand\thebibliography[1]{
  \OLDthebibliography{#1}
  \setlength{\parskip}{0pt}
  \setlength{\itemsep}{0pt plus 0.3ex}
}

\title{Studying the gluon TMDs with $J/\psi$- and $\Upsilon$-pair production at the LHC}

\ShortTitle{Studying the gluon TMDs with $J/\psi$- and $\Upsilon$-pair production at the LHC}

\author{\speaker{Florent Scarpa}$^{\;ab}$, Dani\"el Boer$^{\,a}$,
Miguel G. Echevarria$^{\,c}$, Jean-Philippe Lansberg$^{\,b}$, Cristian Pisano$^{\,d}$, and Marc Schlegel$^{\,e}$\\
\llap{$^a$}Van Swinderen Institute for Particle Physics and Gravity, University of Groningen, Nijenborgh 4, 9747 AG Groningen, The Netherlands\\
\llap{$^b$}IPNO, CNRS-IN2P3, Univ. Paris-Sud, Universit\'e Paris-Saclay, 91406 Orsay Cedex, France\\
\llap{$^c$}Istituto Nazionale di Fisica Nucleare, Sezione di Pavia, via Bassi 6, 27100 Pavia, Italy\\
\llap{$^d$}Dipartimento di Fisica, Universit\`a di Cagliari, and INFN, Sezione di Cagliari
    Cittadella Universitaria, I-09042 Monserrato (CA), Italy\\
\llap{$^e$}Department of Physics, New Mexico State University, Las Cruces, NM 88003, USA\\
E-mail: \email{scarpaflorent@ipno.in2p3.fr}, \email{d.boer@rug.nl},
\email{mgechevarria@pv.infn.it}, \email{Jean-Philippe.Lansberg@in2p3.fr}, \email{cristian.pisano@ca.infn.it}, \email{schlegel@nmsu.edu}}

\abstract{We report on how $J/\psi$- and $\Upsilon$-pair production are promising processes to access the polarised and unpolarised gluon TMDs at the LHC. We present the formalism used, as well as resulting observables that could be extracted from data.}

\FullConference{XXVII International Workshop on Deep-Inelastic Scattering and Related Subjects - DIS2019\\
		8-12 April, 2019\\
		Torino, Italy}

\begin{document}

\vspace{-3mm}
\section{Introduction}\vspace{-3mm}

Transverse-Momentum Dependent (TMD) factorisation allows one to study the correlations between the parton spin and its transverse momentum, inside polarised and unpolarised hadrons \cite{Ralston:1979ys,Sivers:1989cc,Tangerman:1994eh}. These correlations can manifest themselves through azimuthal modulations of the cross-sections for various hadronic processes. At the LHC, the gluon density inside the proton is much higher than the quark ones, making gluon fusion the main channel of heavy-flavoured hadron production. Quarkonium-pair production is a particularly interesting process for the study of gluon TMDs. The quarkonia are dominantely produced via colour-singlet transitions~\cite{Lansberg:2014swa,Lansberg:2013qka,Lansberg:2019fgm,Lansberg:2019adr}. The observed final state is thus colourless  which is a necessary condition not to break TMD factorisation~\cite{Collins:2007nk,Rogers:2010dm}.
Moreover, a two-particle final state allows for large individual momenta for each quarkonium adding up to a small transverse momentum of the pair, a requirement for the use of TMD factorisation.

$J/\psi$- and $\Upsilon$- pair production has already been the object of several experimental studies at the LHC and the Tevatron \cite{Khachatryan:2016ydm,Aaij:2011yc,Abazov:2014qba,Aaboud:2016fzt,Aaij:2016bqq,Khachatryan:2014iia}. This allowed to perform a first fit of the width of a Gaussian-modelled TMD using the normalised transverse-momentum-spectrum of di-$J/\psi$ production at LHCb hinting at the presence of QCD evolution effects~\cite{Lansberg:2017dzg}. More observables could be extracted from data already acquired and to come and this motivates us advancing the theory description of such processes by including TMD QCD evolution as we report on here along the lines of \cite{Scarpa:to_appear}.

\vspace{-3mm}
\section{Quarkonium-pair production within TMD factorisation}\vspace{-3mm}

TMD factorisation extends collinear factorisation by accounting for the parton transverse momenta. The hard-scattering amplitude is factorised into a short-distance coefficient and parton correlators that contain the dependence on the partonic transverse momentum. The gluon correlator for an unpolarised proton can be parametrised in terms of two independent TMDs \cite{Mulders:2000sh,Meissner:2007rx}. 
The first distribution is that of unpolarised gluons $f_1^{\,g}(x,\kT^2)$, the second one is that of linearly polarised gluons $h_1^{\perp\,g}(x,\kT^2)$. The generic structure of the TMD cross-section for quarkonium-pair production from gluon fusion is similar to that of quarkonium-photon production~\cite{Dunnen:2014eta,Lansberg:2017tlc} and reads \cite{Lansberg:2017dzg}:
\begin{align}\label{eq:crosssection} 
&\frac{\d\sigma}{\d M_{\Q\Q} \d Y_{\Q\Q} \d^2 \qT \d \Omega} 
  =\frac{\sqrt{M_{\Q\Q}^2 - 4 M_\mc{Q}^2}}{(2\pi)^2 8 s\, M_{\Q\Q}^2}
  \Bigg\{	F_1\ \mc{C} \Big[f_1^gf_1^g\Big]  + 
  F_2\ \mc{C} \Big[w_2h_1^{\perp g}h_1^{\perp g}\Big]\nn \\+ &
\cos2\phi_{\CS} \ \Bigg(F_3 \ \mc{C} \Big[w_3 f_1^g h_1^{\perp g}\Big]  + F'_3\ \mc{C} \Big[w'_3 h_1^{\perp g} f_1^g \Big]\Bigg)  +  \cos 4\phi_{\CS} \ F_4\ \mc{C}\! \left[w_4 h_1^{\perp g}h_1^{\perp g}\right]\!\Bigg \}\,,
\end{align}	
with 
$\mathcal{C}[w\, f\, g](x_{1,2},\qT) \equiv
\int\!\! \d^{2}\koneT\!\! \int\!\! \d^{2}\ktwoT\,
  \delta^{2}(\koneT+\ktwoT-\bm \qT)\, 
  w(\koneT,\ktwoT)\, 
  f(x_1,\koneT^{2})\, 
  g(x_2,\ktwoT^{2})
$
(the $w_i$ are called TMD weights), 
$\d\Omega=\d\!\cos\theta_{\CS}\d\phi_{\CS}$, $\{\theta_{\CS},\phi_{\CS}\}$ being the Collins-Soper (CS) angles \cite{Collins:1977iv}, $Y_{\Q\Q}$  is the pair rapidity and $s = (P_1 + P_2)^2$.
$\qT$ and $Y_{\Q\Q}$ are considered in the hadron c.m.s.  The $F_i$ coefficients are functions of  $\theta_{\CS}$ and the invariant mass of the pair, denoted $M_{\Q\Q}$.

 The weights in Eq.\ \eqref{eq:crosssection} are identical for all gluon-fusion processes in unpolarised proton collisions and can be found in \cite{Lansberg:2017tlc}. Having at hand the hard-scattering coefficients, one can extract the TMD convolutions from measurements of the cross-section azimuthal modulations. To do so, one defines $\cos(n\phi_{\CS})$-weighted differential cross-sections, integrated over $\phi_{\CS}$ and normalized by their azimuthally-independent component:\vspace{-3mm}
\begin{equation}
\!\!\!\!\langle  \cos(n\phi_{\CS}) \rangle =
\frac{\displaystyle \int \!\!d\phi_{\CS} \cos(n\phi_{\CS})\,  \frac{\d\sigma}{\d M_{\Q\Q} \d Y_{\Q\Q} \d^2 \qT \d \Omega}}
{\displaystyle\!\!\int \!\!d\phi_{\CS} \frac{\d\sigma}{\d M_{\Q\Q} \d Y_{\Q\Q} \d^2 \qT \d \Omega}}\, .
\end{equation}

The TMD QCD evolution introduces a dependence of the TMDs on two scales: a renormalization scale $\mu$ and a rapidity scale $\zeta$ \cite{Collins:2011ca,Echevarria:2012pw,Echevarria:2014rua,Echevarria:2015uaa}.
The evolution is most easily treated in the impact-parameter space, $b_T$ being the conjugate variable to $k_T$. 
TMDs are first computed at the scale $\mu\sim\sqrt{\zeta}\sim\mu_b=b_0/b_T$ (with $b_0=2e^{-\gamma_E}$) to 
minimise the logarithms of $\mu b_T$ and $\zeta b_T^2$, and then evolved up to $\mu\sim\sqrt{\zeta}\sim M_{\Q\Q}$, the natural scale of the hard part. 
Solving the renormalisation group equation and the Collins-Soper equation one can show that such evolution introduces a Sudakov factor $S_A$ in the convolution. 
This factor can be perturbatively evaluated for small values of $b_T$, and is given at next-to-leading-logarithmic accuracy by \cite{Echevarria:2015uaa}:
\begin{align}\label{eq:Sa}
S_A(b_T;\zeta,\mu)&= 
\frac{C_A}{\pi}\int_{\mu_b^2}^{\mu^2}\!\frac{d\bar\mu^2}{\bar\mu^2}
\alpha_s(\bar\mu^2)\left[\Bigg( 
1 
+
\frac{\alpha_s(\bar\mu^2)}{4\pi}\;\frac{67-3\pi^2-20T_fn_f}{9} \Bigg)
\ln\left(\frac{\zeta}{\bar\mu^2}\right)-\frac{11-2n_f/C_A}{6}\right]
\,.
\end{align}

\vspace{-3mm}

In order to perform the Fourier transform, one needs to integrate over all $b_T$ values inside the convolution while paying attention to the validity range of Eq. \eqref{eq:Sa}. Indeed, large $b_T$ values correspond to the non-perturbative region, and too small $b_T$ values would correspond to $\mu_b>M_{\Q\Q}$ for which the expression is not valid anymore. The TMD scale $\mu_b(b_T)$ is thus replaced by $\mu_b\Big(b_T^*\big(b_c(b_T)\big)\Big)$ following \cite{Collins:2016hqq}
such that $\mu_b$ is bound to lie between $b_0/b_{T_{\max}}$ and $M_{\Q\Q}$. 

As far as the TMD expressions are concerned, they can be pertubatively computed at leading order in $\alpha_s$ as
\begin{align}\label{f1pert}
\tilde{f}_1^{\,g}(x,b_T^{*\, 2};\zeta,\mu) &= f_{g/P}(x;\mu)+\mathcal{O}(\alpha_s),\\
\tilde{h}_1^{\,\perp g}(x,b_T^{*\, 2};\zeta,\mu)  &= 
-\frac{\alpha_s(\mu)}{\pi}\int_x^1\!\!\frac{d\hat{x}}{\hat{x}}\left(\frac{\hat{x}}{x}-1\right)\left( C_A\, f_{g/P}(\hat{x};\mu)+C_F\!\!\sum_{i=q,\bar q}f_{i/P}(\hat{x};\mu)\right) + \mathcal{O}(\alpha_s^2)\,.
\end{align}
As $h_1^{\perp\, g}$ describes the correlation between the gluon polarisation and its transverse momentum $k_T$ inside the unpolarised proton, it requires a helicity flip and thus an additional gluon emission. Consequently, its perturbative expansion starts at $\mathcal{O}(\alpha_s)$~\cite{Sun:2011iw}. Such a reasoning only applies for $b_T$ in the perturbative region, outside of which $h_1^{\perp\, g}$ is not computable. 

One still needs to describe the large-$b_T$ behaviour of both the initial TMDs and the perturbative Sudakov factor. The deviation between the components perturbatively evaluated at $b_T^*$ and their actual value is encoded in what is called a nonperturbative Sudakov factor $S_{{\rm NP}}$. 
Including all these ingredients, one gets the expressions for the different TMD convolutions in $b_T$-space to be:
\begin{align}
\label{convs_bt}
&\mathcal{C}\Big[f_1^{\, g}f_1^{\, g}\Big] \!=\! 
\int_0^\infty\!\! \frac{db_T^2}{4\pi}\, J_0(b_T q_T)\, 
e^{-S_A(b_T^*;M_{\Q\Q}^2,M_{\Q\Q}) - S_{{\rm NP}}(b_c)}\,
\tilde{f}_1^{\,g}(x_1,b_T^{*\, 2};\mu_b^2,\mu_b)\, 
\tilde{f}_1^{\,g}(x_2,b_T^{*\, 2};\mu_b^2,\mu_b)
\hspace*{-1cm} ~
\\
&\mathcal{C}\!\Big[w_2\,h_1^{\perp\, g}h_1^{\perp\, g}\Big] \!=\! 
\int_0^\infty\!\! \frac{db_T^2}{4\pi}\, J_0(b_T q_T)\, 
e^{-S_A(b_T^*;M_{\Q\Q}^2,M_{\Q\Q}) - S_{{\rm NP}}(b_c)}\,
\tilde{h}_1^{\perp\, g}(x_1,b_T^{*\, 2};\mu_b^2,\mu_b)\, 
\tilde{h}_1^{\perp\, g}(x_2,b_T^{*\, 2};\mu_b^2,\mu_b) 
\,,
\nonumber\\
&\mathcal{C}\!\Big[w_3\,f_1^{\, g}h_1^{\perp\, g}\Big] \!=\!  
\int_0^\infty\!\! \frac{db_T^2}{4\pi}\, J_2(b_T q_T)\, 
e^{-S_A(b_T^*;M_{\Q\Q}^2,M_{\Q\Q}) - S_{{\rm NP}}(b_c)}\,
\tilde{f}_1^{\,g}(x_1,b_T^{*\, 2};\mu_b^2,\mu_b)\, 
\tilde{h}_1^{\perp\, g}(x_2,b_T^{*\, 2};\mu_b^2,\mu_b)
\,, \nonumber
\\
&\mathcal{C}\!\Big[w_4\,h_1^{\perp\, g}h_1^{\perp\, g}\Big] \!=\!  
\int_0^\infty\!\! \frac{db_T^2}{4\pi}\, J_4(b_T q_T)\, 
e^{-S_A(b_T^*;M_{\Q\Q}^2,M_{\Q\Q}) - S_{{\rm NP}}(b_c)}\,
\tilde{h}_1^{\perp\, g}(x_1,b_T^{*\, 2};\mu_b^2,\mu_b)\, \tilde{h}_1^{\perp\, g}(x_2,b_T^{*\, 2};\mu_b^2,\mu_b)
\,.\nonumber
\end{align}

This $S_{{\rm NP}}$ factor is unknown.
We have used~\cite{Scarpa:to_appear} a very simple form for it in order to estimate its impact over TMD observables, namely
$S_{{\rm NP}}\big(b_c(b_T)\big)=A \ln\Big(\frac{M_{\Q\Q}}{Q_{\rm NP}}\Big)\, b_c^2(b_T)$ with 
$Q_{\rm NP}=1~{\rm GeV}$
where $Q_{\rm NP}$ is a reference scale and $A$ the only free parameter in our model that parametrises the width of the $b_T$-Gaussian. We have performed computations for values of the parameter $A$ equal to 0.64, 0.16 and 0.04 GeV$^2$, which respectively correspond to a factor $e^{-S_{{\rm NP}}}$ becoming smaller than 10$^{-3}$ at $b_{T\lim}$ = 2, 4 and 8 GeV$^{-1}$. 

\vspace{-3mm}

\section{Results for $J/\psi$- and $\Upsilon$-pair production}\vspace{-3mm}

\begin{figure}[hbt!]
\centering
\subfloat[]{\includegraphics[width=4.75cm]{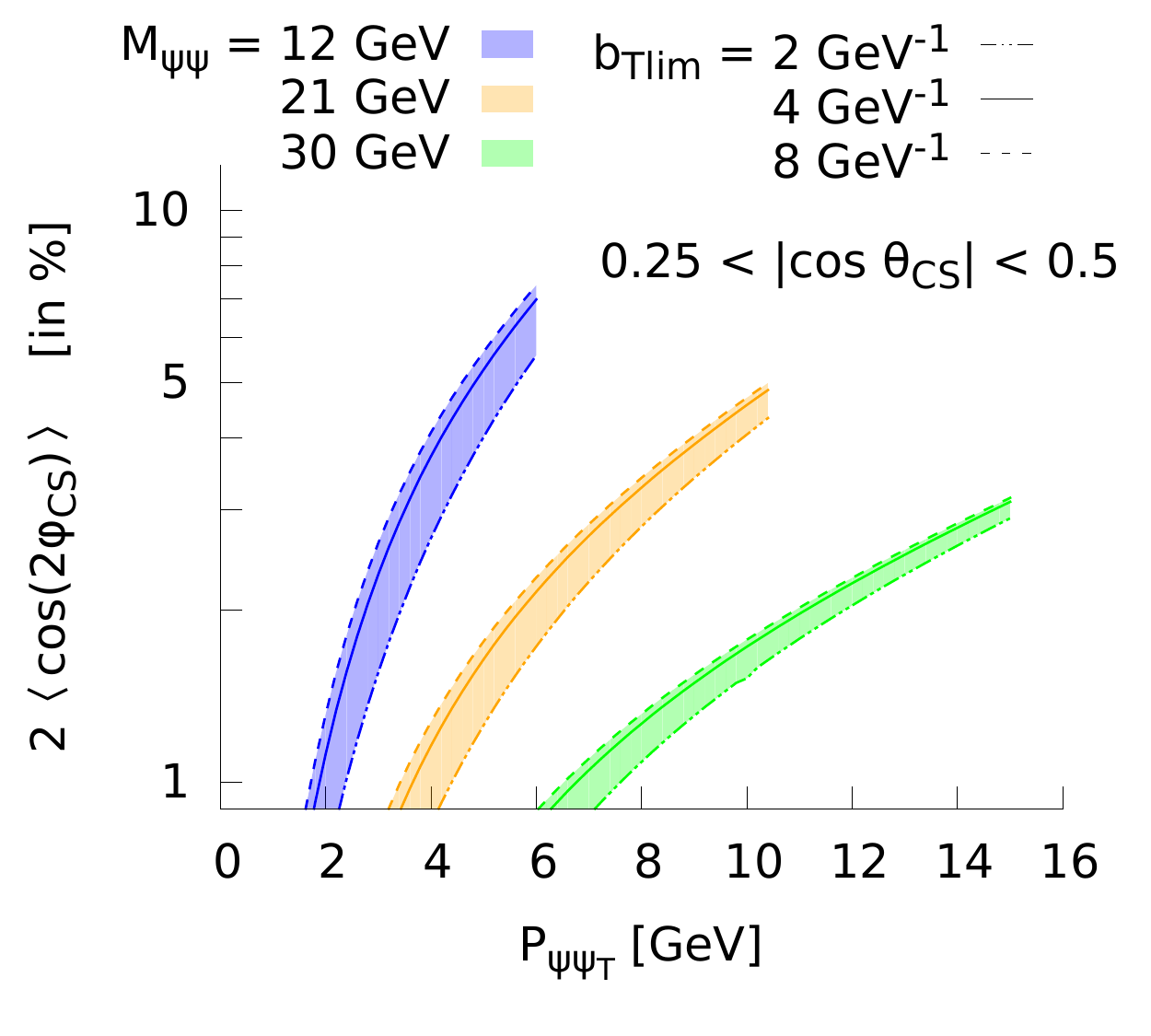}\label{fig:R2_forward}}\hspace{5mm}
\subfloat[]{\includegraphics[width=4.75cm]{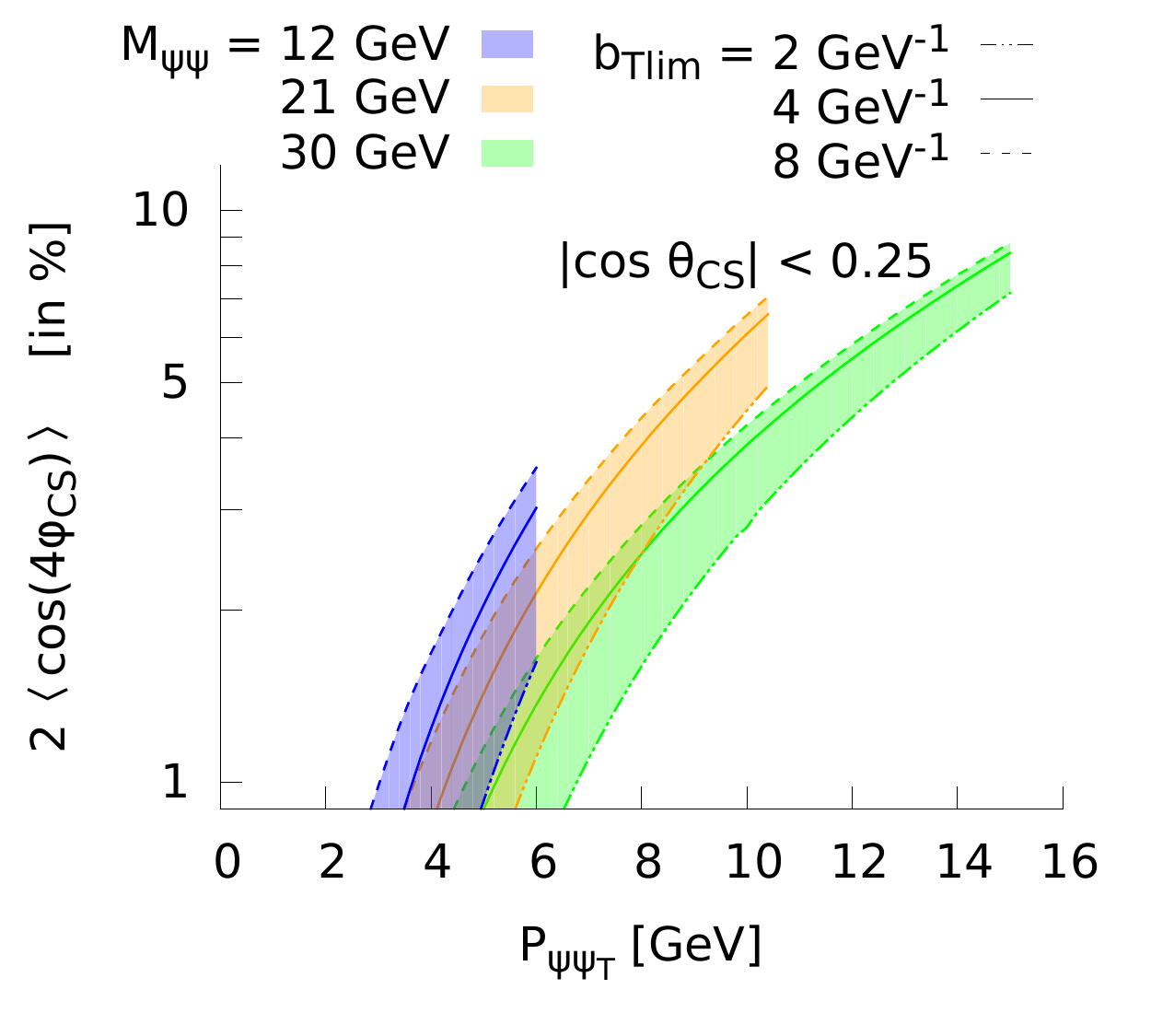}\label{fig:R4_central}}
\caption{
The azimuthal asymmetries for di-$J/\psi$ production as functions of $\qT$. The different plots show $2\langle \cos(2\phi_{CS}) \rangle$ at $0.25<|\cos(\theta_{CS})|<0.5$ (a) and $2\langle \cos(4\phi_{CS}) \rangle$ at $|\cos(\theta_{CS})|<0.25$ (b). 
Results are presented for $M_{\psi\psi}$ = 12, 21 and 30 GeV, and for $b_{T\lim}$ = 2, 4 and 8 GeV$^{-1}$.}
\label{fig:asym}
\subfloat[]{\includegraphics[width=4.75cm]{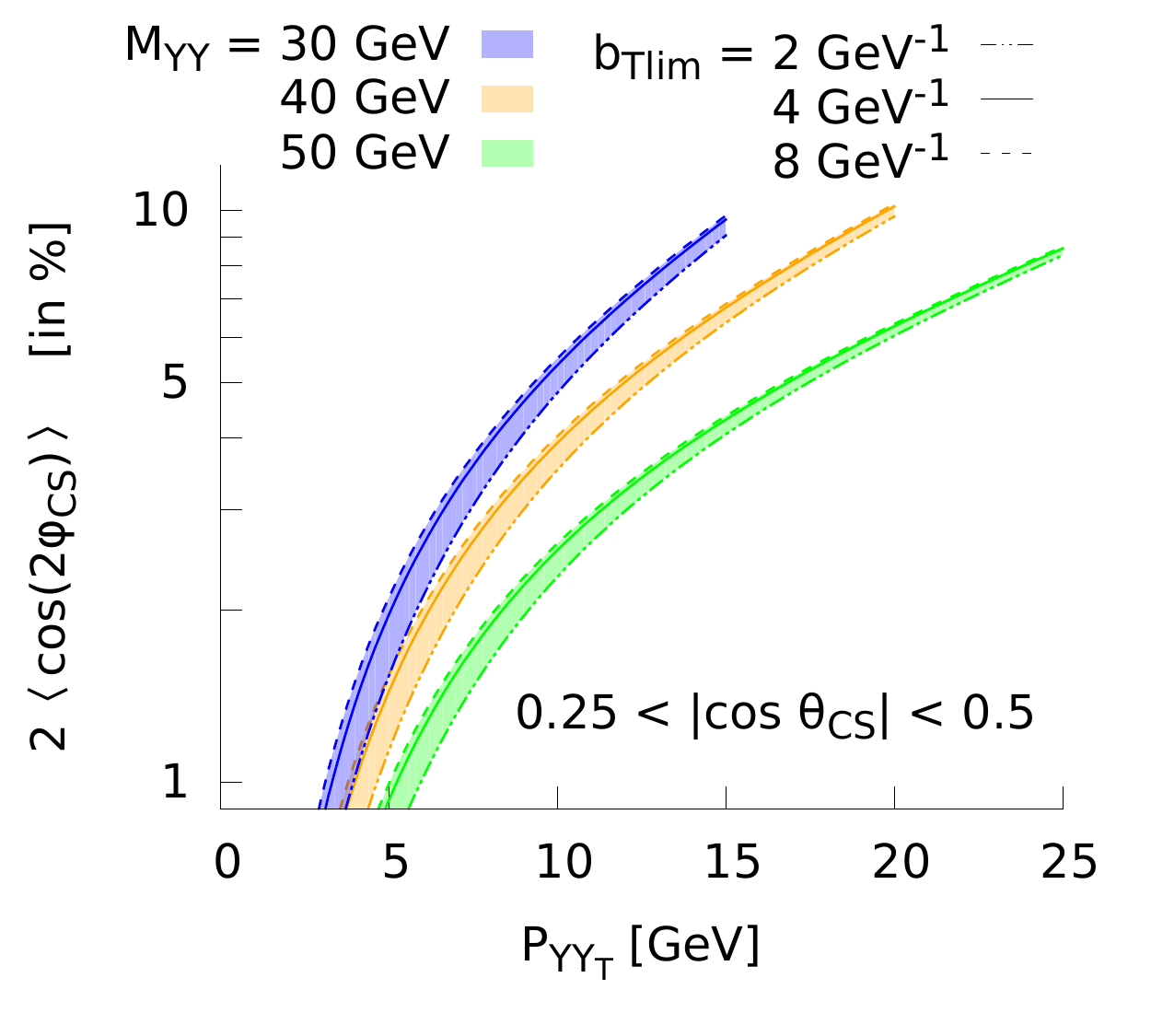}\label{fig:R2_forward_Ups_}}\hspace{5mm}
\subfloat[]{\includegraphics[width=4.75cm]{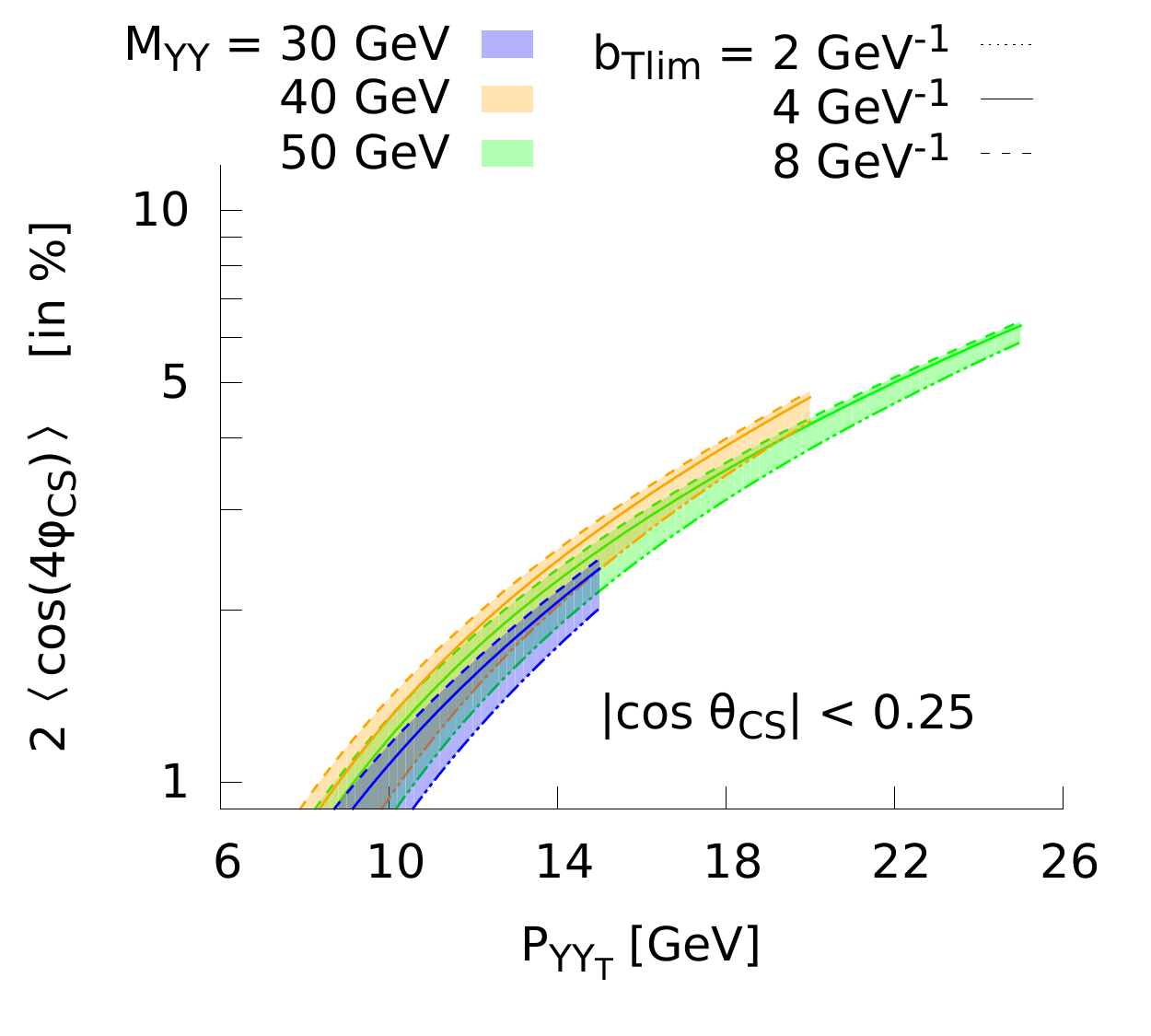}\label{fig:R4_central_Ups}}
\caption{The azimuthal asymmetries for di-$\Upsilon$ production as functions of $\qT$. The different plots show $2\langle \cos(2\phi_{CS}) \rangle$ at $0.25<|\cos(\theta_{CS})|<0.5$ (a) and $2\langle \cos(4\phi_{CS}) \rangle$ at $|\cos(\theta_{CS})|<0.25$ (b). Results are presented for $M_{\Upsilon\Upsilon}$ = 30, 40 and 50 GeV, and for $b_{T\lim}$ = 2, 4 and 8 GeV$^{-1}$.
Results for $M_{\Upsilon\Upsilon}$ = 30 GeV are not included in Fig. 2d as they are below percent level.}
\label{fig:asym_Ups}
\end{figure}

Using Eq. \eqref{convs_bt} and knowing the $F_i$, one can compute the $\cos(2,4\phi)$-asymmetries. In Fig. \ref{fig:asym}, we present the asymmetries for di-$J/\psi$ production as functions of $\qT$.
One can see that the asymmetries grow rapidly with $\qT$, as the convolutions containing $h_1^{\perp\, g}$ have a harder $\qT$-spectrum than $\mathcal{C}\Big[f_1^{\, g}f_1^{\, g}\Big]$. The $M_{\psi\psi}$-dependence is mostly coming from the hard-scattering coefficients. We observe that the uncertainty band generated by the variation of the width of $S_{{\rm NP}}$ is narrower at large scales, where the nonperturbative component of the TMDs is less relevant. The size of the asymmetries can reach up to 10\%.
Fig. \ref{fig:asym_Ups} displays the same asymmetries for di-$\Upsilon$ production, at typically higher energies. At these scales, the $S_{{\rm NP}}$-width uncertainty is quite narrow.

\vspace{-3mm}

\section{Conclusions}\vspace{-3mm}

Quarkonium-pair production offers great opportunities to extract the gluon TMDs inside unpolarised protons. We reported on the formalism used to estimate the size of the related asymmetries, including TMD evolution in the computations. The full analysis can be found in~\cite{Scarpa:to_appear}. We note that more efforts  are needed to obtain rigorous factorisation theorems along the lines of~\cite{Echevarria:2019ynx}. Several asymmetries could be observed at the LHC, both in di-$J/\psi$ and di-$\Upsilon$ production. Accessing these poorly known distributions would improve our understanding of the proton structure, in addition to other TMDs.

\vspace{-3mm}
\section*{Acknowledgements} \vspace{-3mm} 
The work of MS was in part supported within the framework of the TMD Topical Collaboration and
that of FS and JPL by the CNRS-IN2P3 project TMD@NLO. This project has received funding from the European Union's Horizon 2020 research and innovation programme under grant agreement No 824093. MGE is supported by the Marie Sk\l odowska-Curie grant \emph{GlueCore} (grant agreement No. 793896).

\bibliographystyle{utphys}

\providecommand{\href}[2]{#2}\begingroup\raggedright\endgroup

\end{document}